\documentclass[%
 reprint,
 superscriptaddress,
 preprintnumbers,
 bibnotes,
 amsmath,amssymb,
 aps,
 nofootinbib,
longbibliography
]{revtex4-2}
\usepackage{CJK}

 \usepackage[textsize=tiny]{todonotes}

\usepackage[utf8]{inputenc}
\usepackage{graphicx}
\usetikzlibrary{shapes.misc}
\usetikzlibrary{cd}
\usepackage{dcolumn}
\usepackage{bm}
\usepackage{hyperref}
\hypersetup{
    colorlinks=true,
    linkcolor=Nblue,
    filecolor=black,      
    urlcolor=black,
    citecolor=Nblue,
    breaklinks
}

\definecolor{Nblue}{RGB}{0,47,167}   
\definecolor{ku}{RGB}{144,26,30}

\newcommand{\notc}{e}

\newcommand{\dbten}{%
\begin{tikzpicture}[scale=0.1,label distance=-1mm,baseline={([yshift=-.5ex]current bounding box.center)}]
\clip (-0.6,13.5) rectangle (3.5,16.5);
		\node (0) at (0.5, 15.5) {};
		\node (1) at (1.5, 16.5) {};
		\node (4) at (2.5, 15.5) {};
		\node (5) at (0.5, 14.5) {};
		\node (7) at (1.5, 13.5) {};
		\node (8) at (2.5, 14.5) {};
        \node (11) at (5, 15) {};
		\node (12) at (6, 15) {};
		\draw[line width=0.15mm] (0.center) to (4.center);
		\draw[line width=0.15mm] (5.center) to (8.center);
		\draw[line width=0.15mm] (0.center) to (5.center);
		\draw[line width=0.15mm] (1.center) to (7.center);
		\draw[line width=0.15mm] (4.center) to (8.center);
		\draw[line width=0.15mm] (0.5,15.5) to (-0.5, 15.9);
		\draw[line width=0.15mm] (0.5,15.5) to (0.1, 16.5);
		\draw[line width=0.15mm] (0.5,14.5) to (-0.5, 14.1);
		\draw[line width=0.15mm] (0.5,14.5) to (0.1, 13.5);
		\draw[line width=0.15mm] (2.5,15.5) to (3.5, 15.9);
		\draw[line width=0.15mm] (2.5,15.5) to (3.1, 16.5);
		\draw[line width=0.15mm] (2.5,14.5) to (3.5, 14.1);
		\draw[line width=0.15mm] (2.5,14.5) to (3.1, 13.5);
\end{tikzpicture}%
}

\newcommand{\db}{%
\begin{tikzpicture}[scale=0.1,label distance=-1mm,baseline={([yshift=-.5ex]current bounding box.center)}]
\clip (-0.6,13.5) rectangle (3.5,16.5);
		\node (0) at (0.5, 15.5) {};
		\node (1) at (1.5, 15.5) {};
		\node (4) at (2.5, 15.5) {};
		\node (5) at (0.5, 14.5) {};
		\node (7) at (1.5, 14.5) {};
		\node (8) at (2.5, 14.5) {};
        \node (11) at (5, 15) {};
		\node (12) at (6, 15) {};
		\draw[line width=0.15mm] (0.center) to (4.center);
		\draw[line width=0.15mm] (5.center) to (8.center);
		\draw[line width=0.15mm] (0.center) to (5.center);
		\draw[line width=0.15mm] (1.center) to (7.center);
		\draw[line width=0.15mm] (4.center) to (8.center);
		\draw[line width=0.15mm] (0.5,15.5) to (-0.5, 15.9);
		\draw[line width=0.15mm] (0.5,15.5) to (0.1, 16.5);
		\draw[line width=0.15mm] (0.5,14.5) to (-0.5, 14.1);
		\draw[line width=0.15mm] (0.5,14.5) to (0.1, 13.5);
		\draw[line width=0.15mm] (2.5,15.5) to (3.5, 15.9);
		\draw[line width=0.15mm] (2.5,15.5) to (3.1, 16.5);
		\draw[line width=0.15mm] (2.5,14.5) to (3.5, 14.1);
		\draw[line width=0.15mm] (2.5,14.5) to (3.1, 13.5);
		\draw[line width=0.15mm]  (1.5,15.5) to (1.1, 16.5);
		\draw[line width=0.15mm]  (1.5,15.5) to (1.9, 16.5);
		\draw[line width=0.15mm]  (1.5,14.5) to (1.1, 13.5);
		\draw[line width=0.15mm]  (1.5,14.5) to (1.9, 13.5);
\end{tikzpicture}%
}

 \newcommand{\hex}{%
\begin{tikzpicture}[scale=0.1,label distance=-1mm,baseline={([yshift=-.5ex]current bounding box.center)}]
\clip (3.8,13.3) rectangle (7.2,16.7);
       \draw [line width=0.15mm] (4.9,15.4) -- (4.9,14.6);
       \draw [line width=0.15mm] (6.2,15.4) -- (6.2,14.6);
       \draw [line width=0.15mm] (6.2,15.4) -- (5.55,15.8) -- (4.9,15.4);
       \draw [line width=0.15mm] (4.9,14.6)-- (5.55,14.2) --  (6.2,14.6);
       \draw [line width=0.15mm] (5.55,15.8) --(5.2,16.6) ;
       \draw [line width=0.15mm] (5.55,15.8) --(5.9,16.6) ;
       \draw [line width=0.15mm] (5.55,14.2)-- (5.2,13.4);
       \draw [line width=0.15mm] (5.55,14.2)-- (5.9,13.4);
       \draw [line width=0.15mm] (4.9,15.4) -- (3.9, 15.6);
       \draw [line width=0.15mm] (4.9,15.4) -- (4.2, 16.1);
       \draw [line width=0.15mm] (4.9,14.6) -- (4.2, 13.9);
       \draw [line width=0.15mm] (4.9,14.6) -- (3.9, 14.4); 
       \draw [line width=0.15mm] (6.2,15.4) -- (6.9,16.1);
       \draw [line width=0.15mm] (6.2,15.4) -- (7.2,15.6);
       \draw [line width=0.15mm] (6.2,14.6) -- (6.9,13.9);
       \draw [line width=0.15mm] (6.2,14.6) -- (7.2,14.4);
		   \end{tikzpicture}%
		   }

\begin{document}

\begin{CJK*}{UTF8}{gbsn}

\title{Bootstrapping elliptic Feynman integrals using Schubert analysis}
\author{Roger Morales}
 \affiliation{Niels Bohr International Academy, Niels Bohr Institute, Copenhagen University, Blegdamsvej 17, 2100 Copenhagen \O{}, Denmark}
\author{Anne Spiering}
 \affiliation{Niels Bohr International Academy, Niels Bohr Institute, Copenhagen University, Blegdamsvej 17, 2100 Copenhagen \O{}, Denmark}
\author{Matthias Wilhelm}
 \affiliation{Niels Bohr International Academy, Niels Bohr Institute, Copenhagen University, Blegdamsvej 17, 2100 Copenhagen \O{}, Denmark}
\author{Qinglin Yang (杨清霖)}
 \affiliation{%
CAS Key Laboratory of Theoretical Physics, Institute of Theoretical Physics, Chinese Academy of Sciences, Beijing 100190, China}
\author{Chi Zhang (张驰)}
 \affiliation{Niels Bohr International Academy, Niels Bohr Institute, Copenhagen University, Blegdamsvej 17, 2100 Copenhagen \O{}, Denmark}

\begin{abstract}
The symbol bootstrap has proven to be a powerful tool for calculating polylogarithmic Feynman integrals and scattering amplitudes. In this letter, we initiate the symbol bootstrap for elliptic Feynman integrals.
Concretely, we bootstrap the symbol of the twelve-point two-loop double-box integral in four dimensions, which depends on nine dual-conformal cross ratios.
We obtain the symbol alphabet, which contains 100 logarithms as well as 9 simple elliptic integrals, via a Schubert-type analysis, which we equally generalize to the elliptic case.
In particular, we find a compact, one-line formula for the $(2,2)$-coproduct of the result.
\end{abstract}

\maketitle
\end{CJK*}

\section{Introduction}

Within the framework of perturbative Quantum Field Theory, precision predictions are expressed in terms of Feynman integrals, which evaluate to complicated transcendental numbers and functions.

In the last decade, much progress has been made for Feynman integrals, scattering amplitudes, as well as further quantities that belong to the simplest such class of functions, namely multiple polylogarithms (MPLs)~\cite{Chen:1977oja,G91b,Goncharov:1998kja,Remiddi:1999ew,Borwein:1999js,Moch:2001zr}.
This progress is to a large extent due to the excellent understanding we have of these functions, in particular through the so-called symbol and the larger coproduct structure it is part of~\cite{Gonch2,Goncharov:2010jf,Duhr:2011zq,Duhr:2012fh}. 
The symbol allows to decompose MPLs in terms of much simpler symbol letters $\log(\phi_i)$, where $\phi_i$ is a rational or algebraic function of the kinematics.

Among the most powerful techniques we have for MPLs is the so-called symbol bootstrap; see e.g.\ ref.\ \cite{Caron-Huot:2020bkp} for a review.
Since the symbol manifests the identities between MPLs via the known identities of the symbol letters $\log(\phi_i)$, it makes it possible to construct a basis for the space of functions in which a quantity must live. One can then make an ansatz and determine the corresponding coefficients via physical constraints. This idea has been successfully applied to scattering amplitudes 
\cite{Dixon:2011pw,Dixon:2011nj,Dixon:2013eka,Dixon:2014voa,Dixon:2014iba,Drummond:2014ffa,Dixon:2015iva,Caron-Huot:2016owq,Dixon:2016apl,Dixon:2016nkn,Drummond:2018caf,Caron-Huot:2019vjl,Dixon:2020cnr}, 
form factors \cite{Brandhuber:2012vm,Dixon:2020bbt,Guo:2021bym,Dixon:2022rse,Dixon:2022xqh}, soft anomalous dimensions \cite{Li:2016ctv,Almelid:2017qju} and various individual Feynman integrals \cite{Henn:2018cdp,He:2021eec}.
A crucial ingredient for the symbol bootstrap, though, is a good guess for the set of symbol letters, called symbol alphabet. In a growing number of cases, it can be obtained via cluster algebras and tropical Gra\ss{}mannians \cite{Golden:2013xva,Golden:2014pua,Golden:2014xqa,Golden:2014xqf,Drummond:2017ssj,Bourjaily:2018aeq,Drummond:2018dfd,Golden:2018gtk,Drummond:2018caf,Golden:2019kks,Golden:2021ggj,Drummond:2019qjk,Drummond:2019cxm,Arkani-Hamed:2019rds,Henke:2019hve,Drummond:2020kqg,Mago:2020kmp,Chicherin:2020umh,Mago:2020nuv,Herderschee:2021dez,He:2021esx,Mago:2021luw,Henke:2021ity,Ren:2021ztg,Papathanasiou:2022lan} 
as well as, more recently, a Schubert analysis \cite{nima,Yang:2022gko,He:2022tph}.
  
  However, also more complicated classes of functions than MPLs occur in QFT in general and Feynman integrals in particular, see ref.\ \cite{Bourjaily:2022bwx} for a review. The simplest of these are elliptic generalizations of multiple polylogarithms (eMPLs), for which there has been much recent progress \cite{Laporta:2004rb,MullerStach:2012az,brown2011multiple,Bloch:2013tra,Adams:2013nia,Adams:2014vja,Adams:2015gva,Adams:2015ydq,Adams:2016xah,Adams:2017ejb,Adams:2017tga,Bogner:2017vim,Broedel:2017kkb,Broedel:2017siw,Adams:2018yfj,Broedel:2018iwv,Broedel:2018qkq,Honemann:2018mrb,Bogner:2019lfa,Broedel:2019hyg,Duhr:2019rrs,Walden:2020odh,Weinzierl:2020fyx,Giroux:2022wav}.
Specifically, a symbol has been defined for eMPLs \cite{Broedel:2018iwv}, the identities between elliptic symbol letters $\Omega^{(j)}(\tilde\phi_i)$ were understood \cite{Wilhelm:2022wow} and the symbol of the first elliptic Feynman integrals was studied, revealing surprisingly simple structures \cite{Kristensson:2021ani,Wilhelm:2022wow}.
  
\begin{figure}[tb]
\centering
\begin{tikzpicture}[label distance=-1mm]
		\node[label=left:$10$] (0) at (0, 15.5) {};
		\node[label=above:$1$] at (1.7, 16) {};
		\node[label=above:$12$] at (1.3, 16) {};
		\node[label=above:$11$] (2) at (0.5, 16) {};
		\node[label=above:$2$] (3) at (2.5, 16) {};
		\node[label=right:$3$] (4) at (3.0, 15.5) {};
		\node[label=below:$8$] (5) at (0.5, 14) {};
		\node[label=left:$9$] (6) at (0, 14.5) {};
		\node[label=below:$6$] at (1.7, 14) {};
		\node[label=below:$7$] at (1.3, 14) {};
		\node[label=right:$4$] (8) at (3, 14.5) {};
		\node[label=below:$5$] (9) at (2.5, 14) {};
        \node(11) at (5, 15) {};
		\node (12) at (6, 15) {};
		\draw[line width=0.15mm] (0.center) to (4.center);
		\draw[line width=0.15mm] (6.center) to (8.center);
		\draw[line width=0.15mm] (2.center) to (5.center);
		\draw[line width=0.15mm] (3.center) to (9.center);
		\draw[line width=0.15mm] (1.5,15.5) -- (1.5,14.5);
		\draw[line width=0.15mm] (1.5,15.5) -- (1.35,16);
		\draw[line width=0.15mm] (1.5,15.5) -- (1.65,16);
		\draw[line width=0.15mm] (1.5,14.5) -- (1.35,14);
		\draw[line width=0.15mm] (1.5,14.5) -- (1.65,14);				
        \node[label=left:\textcolor{blue!50}{$x_5$}] (10) at (0.2, 15) {};
        \node[label=above:\textcolor{blue!50}{$x_6$}] (13) at (1.0, 15.8) {};
        \node[label=above:\textcolor{blue!50}{$x_1$}] (14) at (2.0, 15.8) {};
        \node[label=right:\textcolor{blue!50}{$x_2$}] (15) at (2.8, 15) {};
        \node[label=below:\textcolor{blue!50}{$x_3$}] (16) at (2.0, 14.3) {};
        \node[label=below:\textcolor{blue!50}{$x_4$}] (17) at (1.0, 14.3) {};
		 \draw[very thick,blue!50] (10.center) to (15.center);
		 \draw[very thick,blue!50] (13.center) to (17.center);
		 \draw[very thick,blue!50] (14.center) to (16.center);
       \draw [line width=0.15mm] (4.9,15.4) -- (4.9,14.6);
       \draw [line width=0.15mm] (6.2,15.4) -- (6.2,14.6);
       \draw [line width=0.15mm] (6.2,15.4) -- (5.55,15.8) -- (4.9,15.4);
       \draw [line width=0.15mm] (4.9,14.6)-- (5.55,14.2) --  (6.2,14.6);
       \draw [line width=0.15mm] (5.55,15.8) --(5.4,16.1) ;
       \draw [line width=0.15mm] (5.55,15.8) --(5.7,16.1) ;
       \draw [line width=0.15mm] (5.55,14.2)-- (5.4,13.9);
       \draw [line width=0.15mm] (5.55,14.2)-- (5.7,13.9);
       \draw [line width=0.15mm] (4.9,15.4) -- (4.6, 15.7);
       \draw [line width=0.15mm] (4.9,15.4) -- (4.5, 15.5);
       \draw [line width=0.15mm] (4.9,14.6) -- (4.6, 14.3);
       \draw [line width=0.15mm] (4.9,14.6) -- (4.5, 14.5); 
       \draw [line width=0.15mm] (6.2,15.4) -- (6.5,15.7);
       \draw [line width=0.15mm] (6.2,15.4) -- (6.6,15.5);
       \draw [line width=0.15mm] (6.2,14.6) -- (6.5,14.3);
       \draw [line width=0.15mm] (6.2,14.6) -- (6.6,14.5); 
       \node at (5.35,16.3) {12};
       \node at (5.75,16.3) {1};
       \node at (5.35,13.7) {7}; 
       \node at (5.75,13.7) {6}; 
       \node at  (6.60,15.85) {2};
       \node at  (6.75,15.45) {3};
       \node at  (6.75,14.55) {4};
       \node at  (6.60,14.15) {5};
       \node at  (4.50,14.15) {8};
       \node at  (4.35,14.55) {9};
       \node at  (4.25,15.5) {10};
       \node at  (4.45,15.85) {11};
\draw[very thick,blue!50] (5.55,15) -- (5.1,15.8) ;
\draw[very thick,blue!50] (5.55,15) -- (6.0,15.8) ;
\draw[very thick,blue!50] (5.55,15) -- (4.7,15) ;
\draw[very thick,blue!50] (5.55,15) -- (6.4,15) ;
\draw[very thick,blue!50] (5.55,15) -- (5.1,14.2) ;
\draw[very thick,blue!50] (5.55,15) -- (6.0,14.2) ;
\node at  (5,15.95) {\textcolor{blue!50}{$x_6$}};
\node at  (6.05,15.95) {\textcolor{blue!50}{$x_1$}};
\node at (6.65,15) {\textcolor{blue!50}{$x_2$}};
\node at (4.4,15) {\textcolor{blue!50}{$x_5$}};
\node at (6.15,14.05) {\textcolor{blue!50}{$x_3$}};
\node at (5,14.05) {\textcolor{blue!50}{$x_4$}};
\end{tikzpicture} 
\caption{The twelve-point elliptic double box and the related hexagon, as well as their dual graphs. The dual momenta are defined via $x_{i+1}-x_{i}=p_{2i}+p_{2i+1}$.}
\label{fig: integral 12pt}
\end{figure}
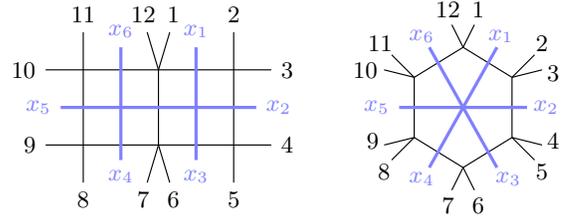

  In this letter, we initiate the symbol bootstrap for elliptic Feynman integrals.
  Concretely, we calculate the twelve-point two-loop double-box integral with massless internal propagators in four spacetime dimensions, see Fig.\ \ref{fig: integral 12pt}.
  This diagram contributes in particular to scattering amplitudes in the maximally supersymmetric Yang-Mills ($\mathcal{N}=4$ sYM) theory \cite{Bourjaily:2015jna} and, through its dual graph, to correlation functions in that theory as well as its conformal fishnet limit \cite{Gurdogan:2015csr,Sieg:2016vap,Grabner:2017pgm}.
  Our bootstrap is based on the structures that were observed in the ten-point double-box integral \cite{Kristensson:2021ani} -- in particular the symbol prime \cite{Wilhelm:2022wow} -- as well as on generalizing the Schubert analysis to the elliptic case.

\section{Setup}
\label{sec: setup}
  
  We consider the twelve-point double-box integral 
  \begin{equation} \label{db_integrand}
  \begin{aligned}
  I_{\db} =&\int \frac{d^4 x_{0}\, d^4 x_{0'} \quad x_{14}^{2}x_{25}^{2}x_{36}^{2}}{x_{10}^{2}x_{20}^{2}x_{30}^{2}x_{00'}^{2}x_{40'}^{2}x_{50'}^{2}x_{60'}^{2}}\,,
   \end{aligned}
\end{equation}
with 
$x_{ij}^{2}\equiv(x_{i}-x_{j})^{2}$. Note that no pair of the six external dual points $x_{i}$ is light-like separated, i.e.\ $x_{ij}^{2}\neq 0$ for all $i\neq j$, and the numerator is chosen to render the diagram dual conformal invariant.
The double-box integral~\eqref{db_integrand} depends on nine independent dual-conformal cross ratios
\begin{equation}
	\chi_{ab}=\chi_{ba}=x_{b\,a-1;a\,b-1} \quad\text{with}\quad
	  x_{ij;kl}=\frac{x_{ij}^2 x_{kl}^2}{x_{ik}^2 x_{jl}^2}\,,
\end{equation}
where  
$a,b$ are nonadjacent in the cycle $\{1,\ldots ,6\}$.
Moreover, it satisfies a first-order differential equation relating it to the one-loop hexagon integral in six dimensions \cite{Paulos:2012nu,Nandan:2013ip}:
\begin{equation}
\label{eq: differential equation}
\partial_{\chi_{14}} I_{\db} = \frac{1}{\sqrt{-\Delta_6}}I_{\hex}\,,
\end{equation}
where the normalized six-point Gram determinant $\Delta_6= \det (x_{ij}^{2})/(x_{14}^{2}x_{25}^{2}x_{36}^{2})^{2}$ is a cubic polynomial in $\chi_{14}$, showing that the double-box integral is elliptic.  

We are interested in the singularity structure of the integral \eqref{db_integrand}, i.e.\ its \emph{symbol}~\cite{Goncharov:2010jf,BrownNotes,Broedel:2018iwv}, which can be obtained by taking the total differential recursively,
\begin{equation}
	d I=\sum_i I_i \,d A_i \:\: \Rightarrow \:\: \mathcal{S}(I)=\sum_i \mathcal{S}(I_i) \otimes  A_i\,,
   \end{equation}
where $I$, $I_{i}$, and the \emph{symbol letters} $A_{i}$ are $n$-, $(n{-1})\text{-,}$ and 1-fold integrals, respectively.
It was computed in ref.\ \cite{Kristensson:2021ani} and further indicated in ref.\ \cite{Wilhelm:2022wow} that the symbol of the  ten-point double-box integral, given by the limit $x_{16}^2\to0$ and $x_{34}^2\to0$ of eq.\ \eqref{db_integrand}, respects the following simple structure:
\begin{multline}
\label{eq: symbol structure}
\mathcal{S}\left( \frac{2\pi i}{\omega_1} I_{\dbten}\right)=  \sum_{ikl} C^{ikl}\log({\phi_{k}})\otimes\log({\phi_{l}})
\\\otimes \! \Biggl[\sum_j \log ({\phi_{ij}}) \otimes \left(2\pi i\,w_{c_j}\right) + {\bm\Omega}_{i}\otimes (2 \pi i\,{\tau})   \Biggr]
 , 
\end{multline}
where $\omega_1$ and $\omega_2$ are the periods of the elliptic curve, with modular parameter $\tau=\omega_2/\omega_1$, and $C^{ijk}\in\mathbb{Q}$. The elliptic symbol letters in the last entry are 
\begin{equation}
\label{eq:Abel_map}
    w_{c}=\frac{1}{\omega_1}\int_{-\infty}^{c} \frac{dx}{y}\,,
\end{equation} 
with $y^2=-\Delta_6(x)$ defining the elliptic curve.\footnote{Note that we are defining the elliptic curve via a cubic polynomial here, while is was defined via a quartic polynomial in ref.\ \cite{Kristensson:2021ani}. The two curves can be birationally mapped to each other, though.}
Using the symbol prime \cite{Wilhelm:2022wow}, the remaining elliptic letters $\bm{\Omega}_{i}$ can be obtained from the previous letters as 
\begin{equation} \label{eq:intrep_symbolprime}
 \bm{\Omega}_{i}=\sum_j \partial_\tau\int_{\gamma}\left(2\pi i\,w_{{c_j}}\right) d\log({ \phi_{ij}})\,,
\end{equation}
where the integration contour $\gamma$ is independent of $\tau$ and will be further discussed in Sec.\ \ref{sec: results}.

It is as-yet-unknown how to evaluate the twelve-point double-box integral in terms of eMPLs and then compute its symbol. 
The main obstacle in applying techniques such as differential equations  or direct integration
is the occurrence of excessive square roots. This can be anticipated from eq.~\eqref{eq: differential equation} as the symbol of the hexagon~\cite{Spradlin:2011wp},
\begin{align}
	\label{eq: hexagon symbol}
	\mathcal{S}(I_{\hex})=\sum_{i<j}\text{Box}_{ij}\otimes \log 
	R_{ij}\,,\quad R_{ij}=\frac{\mathcal{G}_{j}^{i}-\sqrt{-\mathcal{G}_{ij}\mathcal{G}_{}}}{\mathcal{G}_{j}^{i}+\sqrt{-\mathcal{G}_{ij}\mathcal{G}_{}}}\,,
\end{align}
contains square roots of 16 different Gram determinants! 
Here Box$_{ij}$ refers to the symbol of the four-mass box integral 
\begin{equation}
\text{Box}_{ij}=\log v_{ij}\otimes\log \frac{z_{ij}}{\bar z_{ij}}-\log u_{ij}\otimes\log\frac{1-z_{ij}}{1-\bar z_{ij}}\,,
\label{eq:Box}
\end{equation}
with $ u_{ij}=x_{kl;mn}$ and $ v_{ij}=x_{lm;nk}$ for $\{k,l,m,n\}=\{1,\dots,6\}\backslash\{i,j\}$, as well as  $z_{ij}$ and $\bar{z}_{ij}$ being defined by $u_{ij}=z_{ij}\bar{z}_{ij}$ and $v_{ij}=(1-z_{ij})(1-\bar{z}_{ij})$.
Moreover, we introduced the following notation for Gram determinants:
\begin{equation}
	\mathcal{G}_{B}^{A}:=(-1)^{\sum_{c\in\{A,B\}}c}\det x_{ab}^{2}\quad \text{and} \quad  \mathcal{G}_{A}:=\mathcal{G}_{A}^{A}\:,
\end{equation}
with $a\in \{1,\dots,6\}\setminus \{A\}$ and $b\in \{1,\dots,6\}\setminus \{B\}$ where $A,B$ are indices of dual points as in eq.~\eqref{eq: hexagon symbol}; in particular, $\mathcal{G}$ is the six-point Gram determinant.

In this letter, we turn to bootstrapping integral \eqref{db_integrand}, assuming in particular that the structure \eqref{eq: symbol structure} holds also in the twelve-point case.

\section{Symbol letters via a Schubert problem}
\label{sec: schubert}

We now predict the symbol letters required for the bootstrap by using Schubert analysis.
These letters include the logarithmic letters -- in particular those indicated by the symbol of the one-loop hexagon \eqref{eq: hexagon symbol} through eq.~\eqref{eq: differential equation} --
and the elliptic last entries, 
while the complicated letters $\bm{\Omega}_i$ can be constructed from these via eq.~\eqref{eq:intrep_symbolprime}.

Schubert analysis works in twistor space $\mathbb{P}^3$ \cite{Hodges:2009hk,Mason:2009qx}, where to each dual point $x_i^{\alpha\dot{\alpha}}=x_{i}^{\mu}\sigma_{\mu}^{\alpha\dot{\alpha}}$ is associated a line
\begin{equation} \label{eq: line_def}
 (i)= (1,t,x_i^{1\dot{1}}+x_i^{1\dot{2}}t,x_i^{2\dot{1}}+x_i^{2\dot{2}}t)\,,
\end{equation}
where the points are parametrized by $t$.

\begin{figure}[tb]
	\centering
	\begin{tikzpicture}[label distance=-1mm]
			\node (0) at (-0.1, 15.7) {};
			\node[label=above:$\textcolor{blue!50}{(L_1)}$] (1) at (0.5, 17.1) {};
			\node[label=above:$\textcolor{blue!50}{(L_2)}$] (3) at (1.3, 17.1) {};
			\node (4) at (6.9, 15.7) {};
			\node (6) at (-0.1, 14.3) {};
			\node (2) at (0.5, 13.9) {};
			\node (8) at (6.9, 14.3) {};
			\node (21) at (6.9, 15) {};
			\node (22) at (-0.1, 15) {};
			\node (9) at (1.3, 13.9) {};
			\node[label=above:$\textcolor{blue!50}{(M_1)}$] (13) at (3.0, 17.1) {};
			\node (14) at (3.0, 13.9) {};
			\node[label=above:$\textcolor{blue!50}{(M_2)}$] (15) at (3.8, 17.1) {};
			\node (16) at (3.8, 13.9) {};
			\node[label=above:$\textcolor{blue!50}{(N_1)}$] (17) at (5.5, 17.1) {};
			\node (18) at (5.5, 13.9) {};
			\node[label=above:$\textcolor{blue!50}{(N_2)}$] (19) at (6.3, 17.1) {};
			\node (20) at (6.3, 13.9) {};
			\draw[very thick,blue!50] (3.center) to (9.center);
			\draw[very thick,blue!50] (1.center) to (2.center);
			\draw[very thick,blue!50] (13.center) to (14.center);
			\draw[very thick,blue!50] (15.center) to (16.center);
			\draw[very thick,blue!50] (17.center) to (18.center);
			\draw[very thick,blue!50] (19.center) to (20.center);
			\draw[line width=0.15mm] (0.center) to (4.center);
			\draw[line width=0.15mm] (6.center) to (8.center);
			\draw[line width=0.15mm] (21.center) to (22.center);
			\draw[line width=0.15mm] (-0.1,16.4) to (1.9,16.8);
			\draw[line width=0.15mm] (2.4,16.8) to (4.4,16.8);
			\draw[line width=0.15mm] (6.9,16.4) to (4.9,16.8);
			\draw[fill=ku] (0.5,15.7) circle (2pt);
			\draw[fill=ku] (0.5,15) circle (2pt);
			\draw[fill=ku] (0.5,14.3) circle (2pt);
			\draw[fill=ku] (1.3,15.7) circle (2pt);
			\draw[fill=ku] (1.3,15) circle (2pt);
			\draw[fill=ku] (1.3,14.3) circle (2pt);
			\draw[fill=ku] (3.0,15.7) circle (2pt);
			\draw[fill=ku] (3.0,15) circle (2pt);
			\draw[fill=ku] (3.0,14.3) circle (2pt);
			\draw[fill=ku] (3.8,15.7) circle (2pt);
			\draw[fill=ku] (3.8,15) circle (2pt);
			\draw[fill=ku] (3.8,14.3) circle(2pt);
			\draw[fill=ku] (5.5,15.7) circle (2pt);
			\draw[fill=ku] (5.5,15) circle (2pt);
			\draw[fill=ku] (5.5,14.3) circle(2pt);
			\draw[fill=ku] (6.3,15.7) circle (2pt);
			\draw[fill=ku] (6.3,15) circle (2pt);
			\draw[fill=ku] (6.3,14.3) circle(2pt);
			
			\draw[fill=ku] (0.5,16.52) circle (2pt);
			\draw[fill=ku] (1.3,16.68) circle (2pt);
			\draw[fill=ku] (3.0,16.8) circle(2pt);
			\draw[fill=ku] (3.8,16.8) circle (2pt);
			\draw[fill=ku] (5.5,16.68) circle (2pt);
			\draw[fill=ku] (6.3,16.52) circle(2pt);
	
			\draw[rounded corners] (4.8, 13.8) rectangle (7.1, 17.65) {};
			\draw[rounded corners] (2.2, 13.8) rectangle (4.6, 17.65) {};
			\draw[rounded corners] (-0.6,13.8) rectangle (2.0, 17.65) {};
			\node[label=below:$\textcolor{ku}{\alpha_{1}}$] at (0.3,14.3) {};
			\node[label=below:$\textcolor{ku}{\beta_{1}}$] at (0.3,15) {};
			\node[label=below:$\textcolor{ku}{\gamma_{1}}$] at (0.3,15.7) {};
			\node[label=below:$\textcolor{ku}{\delta_{1}}$] at (0.3,16.5) {};
			\node[label=below:$\textcolor{ku}{\alpha_{2}}$] at (1.6,14.3) {};
			\node[label=below:$\textcolor{ku}{\beta_{2}}$] at (1.6,15) {};
			\node[label=below:$\textcolor{ku}{\gamma_{2}}$] at (1.6,15.7) {};
			\node[label=below:$\textcolor{ku}{\delta_{2}}$] at (1.6,16.76) {};
			\node[label=below:$x_j$] at (-0.3,15.2) {};
			\node[label=below:$x_l$] at (-0.3,16.6) {};
			   \node[label=below:$x_k$] at (-0.3,15.9) {};
			\node[label=below:$x_i$] at (-0.3,14.5) {};
	\end{tikzpicture}
	\caption{Schubert problem arising from the mixture of three one-loop Schubert problems. The horizontal lines represent external dual points $x_{i}$ and each pair of vertical lines in the three boxes represents the solution of the corresponding one-loop Schubert problem.}
	\label{fig: Schubert intersection 3 boxes}
	\end{figure}
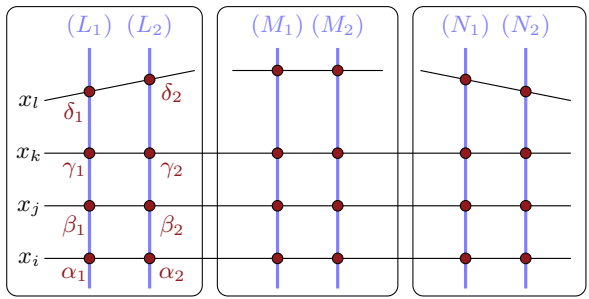

\paragraph{MPL letters from boxes.}
As an introduction to Schubert analysis, let us begin by discussing the one-loop four-mass box integral, whose symbol is given in eq.\ \eqref{eq:Box}. To solve for the one-loop leading singularity of this integral, we send its four propagators 
to zero, i.e.\ $x_{i0}^{2}=0$. 
In momentum twistor space, this is equivalent to looking for a line $(L)$ intersecting all four kinematics lines $(i)$ simultaneously.
There are exactly two solutions $(L_j)_{j=1,2}$ to this so-called Schubert problem, referred to as ``one-loop Schubert problem'' in the following. 
Each of these solutions has four distinct intersections with the four external lines~\cite{ArkaniHamed:2010gh,Bourjaily:2013mma}, $\{\alpha_j,\beta_j,\gamma_j,\delta_j\}_{j=1,2}$; see Fig.\ \ref{fig: Schubert intersection 3 boxes}. According to ref.\ \cite{Yang:2022gko}, one can form four multiplicatively independent cross ratios from these intersections, 
\begin{align}
 z=\frac{(\alpha_1-\beta_1)(\gamma_1-\delta_1)}{(\alpha_1-\gamma_1)(\beta_1-\delta_1)}, \quad
 \bar{z}=(1\to2),
\end{align}
as well as $(1-z)$ and $(1-\bar{z})$.\footnote{Note that while the $\{\alpha_j,\beta_j,\gamma_j,\delta_j\}_{j=1,2}$ depend on the parametrization we chose for the line, the four cross ratios are parametrization independent.}
Taking their products (quotients) we obtain the arguments of the letters for the first (second) entries of the four-mass box symbol \eqref{eq:Box}.

An interesting property of all known amplitudes and Feynman integrals in planar $\mathcal{N}=4$ sYM theory~\cite{Caron-Huot:2020bkp,CaronHuot:2011ky,He:2020vob,He:2020lcu,Kristensson:2021ani,Li:2021bwg,He:2022ujv}, which arguably holds to all loop orders~\cite{Gaiotto:2011dt,He:2021mme}, is that their first two entries satisfy a particular pattern: they form the symbols of $\operatorname{Li}_{2} (1 - x_{ab;cd})$, $\log(x_{ab;cd} ) \log(x_{a'b';c'd'} )$ or four-mass boxes whose symbol letters are $\{z,\bar{z},1{-}z,1{-}\bar{z}\}$ or their degenerations for corresponding one-loop-box sub-diagrams. We assume that the twelve-point double-box integral \eqref{db_integrand} also \emph{follows this pattern}. Since there are $\binom{6}{4}=15$ four-mass box sub-topologies, this gives us $\bm{9}$ candidates for the first entry and $30{+}9=\bm{39}$ candidates for the second entry.

Now we move to the third entries. In ref.\ \cite{Yang:2022gko} it was realised that for certain two-loop planar Feynman integrals, the space of possible symbol letters in the third slot is generated by combining different one-loop Schubert problems and constructing cross ratios from the intersection points on the \emph{external} lines. Here we refine this procedure as follows: 
in all known examples we observe that the required combined one-loop boxes share \emph{three} external lines and thus we assume this to also hold for the twelve-point double box; see Fig.\ \ref{fig: Schubert intersection 3 boxes}.
This requirement in particular guarantees 
that the cross ratios formed on each line are the same.\footnote{This can be proven as follows. The three common external lines  (for generic kinematics) define a quartic surface in $\mathbb{P}^3$, which can be brought into the form $X_1X_4-X_2X_3$ for $[X_1:X_2:X_3:X_4]\in \mathbb{P}^3$ and is thus isomorphic to $\mathbb{P}^1\times \mathbb{P}^1$ via the Segre embedding. The dependence on the external line is contained in the coordinates of the first $\mathbb{P}^1$, while these coordinates are the same for each point on such a line. Thus, these coordinates cancel in the cross ratios, and with them the dependence on the line. 
We are grateful to Cristian Vergu for giving this proof.
}
There are $\binom{6}{3}=20$ such configurations in the double-box integral, each of them giving 9 multiplicatively independent cross ratios. Taking the union of all cross ratios formed in this way, we find $104$ multiplicatively independent letters: $9$ $\chi_{ab}$, $\mathcal{G}_{56}/(x_{13}^{2}x^{2}_{24})^{2}$ and its 14 images under the permutations $S_{6}$ of the external points $x_i$, the $15$ last entries $R
_{ij}$ of the hexagon \eqref{eq: hexagon symbol}, $5$ ratios of $\mathcal{G}_{6}/(x_{13}^{2}x_{24}^{2}x_{35}^{2}x_{25}^{2}x_{14}^{2})$ to its 5 images under $S_{6}$, as well as $60$  algebraic letters
\begin{equation}
	\frac{\mathcal G_{ij}^{ik}-\sqrt{\mathcal G_{ij}\mathcal G_{ik}}}{\mathcal G_{ij}^{ik}+\sqrt{\mathcal G_{ij}\mathcal G_{ik}}}\,.
\end{equation}
Combining them with the 30 ratios $\{z/\bar{z},(1-z)/(1-\bar{z})\}$ from the second entries, we obtain $\bm{134}$ candidate third entries.

\begin{figure}[tb]
	\centering
	\begin{tikzpicture}[label distance=-1mm]
			\node[label=above:$x_1$] (1) at (0.5, 17.1) {};
			\node[label=above:$x_2$] (3) at (1.3, 17.1) {};
			\node (2) at (0.5, 15.9) {};
			\node (9) at (1.3, 15.9) {};
			\node[label=above:$x_3$] (13) at (2.1, 17.1) {};
			\node (14) at (2.1, 15.9) {};
			\node[label=above:$x_4$] (15) at (4.7, 17.1) {};
			\node (16) at (4.7, 15.9) {};
			\node[label=above:$x_5$] (17) at (5.5, 17.1) {};
			\node (18) at (5.5, 15.9) {};
			\node[label=above:$x_6$] (19) at (6.3, 17.1) {};
			\node (20) at (6.3, 15.9) {};
			\draw[line width=0.15mm] (3.center) to (9.center);
			\draw[line width=0.15mm] (1.center) to (2.center);
			\draw[line width=0.15mm] (13.center) to (14.center);
			\draw[line width=0.15mm] (15.center) to (16.center);
			\draw[line width=0.15mm] (17.center) to (18.center);
			\draw[line width=0.15mm] (19.center) to (20.center);
			 \draw[very thick,blue!50] (-0.1,16.4) to (3.9,16.8);
			 \node[label=left:$\textcolor{blue!50}{(0)}$] at (-0.1,16.4) {};
			 \draw[very thick,blue!50] (6.9,16.4) to (2.9,16.8);
			 \node[label=right:$\textcolor{blue!50}{(0')}$] at (6.9,16.4) {};
			 \draw[fill=ku] (0.5,16.46) circle (2pt);
			 \draw[fill=ku] (1.3,16.54) circle (2pt);
			 \draw[fill=ku] (2.1,16.62) circle (2pt);
			 \draw[fill=ku] (4.7,16.62) circle (2pt);
			 \draw[fill=ku] (5.5,16.54) circle (2pt);
			 \draw[fill=ku] (6.3,16.46) circle (2pt);
			 \draw[dotted, ultra thick] plot[smooth, tension=1] coordinates {(3.8,17.6) (3.4,17.2) (3.4,16.2) (3,15.8)};
			 \draw[fill=black] (3.4,16.75) circle (2pt);
	\end{tikzpicture}
	\caption{Two-loop Schubert problem describing the elliptic curve (set of black points) in twistor space \cite{Vergu:2020uur}. If the black point lies on an external line, i.e.\ the elliptic curve intersects the external line in that point, either $(0)$ or $(0')$ solves a one-loop Schubert problem with $x_1,x_2,x_3$ or $x_4,x_5,x_6$, respectively.}
	\label{fig: Schubert elliptic}
	\end{figure}
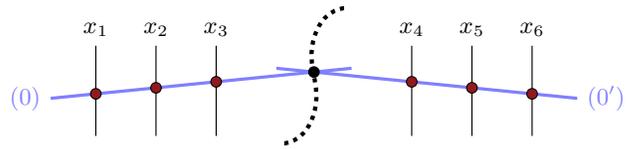

\paragraph{Elliptic Schubert analysis and last entries.}

So far, we have only constructed the arguments of the symbol letters $\log(\phi_i)$ through Schubert analysis, while, as indicated in refs.\ \cite{Kristensson:2021ani,Wilhelm:2022wow}, the counterparts of elliptic letters in MPL cases are logarithms rather than their arguments. However, one can also naturally construct logarithms in the above Schubert analysis; e.g.
\begin{equation}\label{eq: Aomoto}
 \log(z)=(\alpha_1-\delta_1)\int_{\beta_1}^{\gamma_1}\frac{dx}{(x-\alpha_1)(x-\delta_1)}\,.
\end{equation}
This is referred to as Aomoto polylogarithms \cite{Aomoto1982,Goncharov:2009lql,Arkani-Hamed:2017ahv}: two points on the line define the differential form (integrand) and the two other points define the integration range, while the normalization factor $\alpha_{1}-\delta_{1}$ arises from the inverse of the contour integral
\begin{equation} \label{eq: normalization}
	\frac{1}{\alpha_{1}-\delta_{1}}=\frac{1}{2\pi i} \oint_{\lvert x-\alpha_{1}\rvert=\epsilon}\frac{dx}{(x-\alpha_1)(x-\delta_1)} \:,
\end{equation}
which can also be understood as one period of the punctured sphere $\mathbb{C}\setminus\{\alpha_{1},\delta_{1}\}$.

The above construction can be easily generalized to elliptic cases. Concretely,
this amounts to considering a ``two-loop Schubert problem'', corresponding to the leading singularity of the two-loop double-box integral: the lines $(0)$ and $(0')$ intersect each other as well as $\{(1),(2),(3)\}$ and $\{(4),(5),(6)\}$, respectively; see Fig.\ \ref{fig: Schubert elliptic}.
Since these are seven constraints on eight parameters, they define a curve, to which a one-form is naturally associated.
One can easily verify that this is an elliptic curve~\cite{Vergu:2020uur}, and the elliptic generalization of eq.~\eqref{eq: Aomoto} is
\begin{equation}\label{eq: elllast}
\frac{2\pi i}{\omega_1}\int^{\beta}_{\alpha}\frac{dx}{\sqrt{P(x)}}\,.
\end{equation}
Here $dx/\sqrt{P(x)}$ is the differential form for the elliptic curve, with $x$ parametrizing the points on any external line $(i)$. Moreover, $(2\pi i)/\omega_{1}$ is the counterpart of eq.~\eqref{eq: normalization},\footnote{Alternatively, one can choose $-(2\pi i)/\omega_{2}$ as the normalization factor, amounting to the corresponding alternative normalization of the elliptic Feynman integrals; see ref.\ \cite{Wilhelm:2022wow} for further details.} and $\{\alpha,\beta\}$ are intersections on $(i)$ that stem from a one-loop Schubert problem including either $\{(1),(2),(3)\}$ or 
$\{(4),(5),(6)\}$. 
For instance, 
if we stick to the line $(2)$ and choose the upper and lower bounds from $\{\alpha_1,\alpha_2\}$ in Fig.\ \ref{fig: Schubert intersection 3 boxes} with $\{i,j,k,l\}=\{1,2,3,4\}$, the integral gives $w_{\chi_{14}}$, which will be one of our last entries. 
By going through all external lines and possible upper and lower bounds,\footnote{There is some ambiguity in choosing the integration contour in eq.~\eqref{eq: elllast} since the integral is effectively defined on a torus. One can only find all last entries after taking all non-equivalent contours into account. See App.\ \ref{sec: Schubert appendix} for more details.} we find $\bm{9}$ linear independent elliptic integrals which we assume to be the last entries of the twelve-point double-box integral.

Finally, let us remark that these $8$ letters besides $w_{\chi_{14}}$ can also be generated from the differential equation \eqref{eq: differential equation} as the values of $\chi_{14}$ for which the third letters $R_{ij}$ in the hexagon become singular. In this way, we find the overcomplete set of last entries $\{w_{0},w_{\chi_{14}},\tau,w_{d_{k}^{\pm}}\}$, where the kinematic points and their images on the curve read
\begin{align}
\hspace{-0.25cm} d_{k}^{\pm}=& \, \chi_{14} \bigg(1 - \frac{((-1)^{i+k}\mathcal{G}_{ij}^{ik}\pm\sqrt{\mathcal{G}_{ij}\mathcal{G}_{ik}})x_{ij}^{2}x_{ik}^{2}}{2x_{12}^{2}x_{23}^{2}x_{31}^{2}x_{45}^{2}x_{56}^{2}x_{64}^{2}} \bigg), \\
\hspace{-0.25cm} y(d_k^\pm)=& \, \frac{\mathcal{G}_{j}^{i}}{x_{14}^2x_{25}^2x_{36}^2x_{ln}^2x_{km}^2} \sqrt{x_{ln}^2x_{km}^2 (\mathcal{G}_{ij})^{-1}}\bigg|_{\chi_{14}=d_k^\pm}\,.
\label{eq:images_cipm}
\end{align}
Here $i$ and $j$ are defined from the index $k$ by identifying the set $\{i,j,k\}$ with (cyclic permutations of) $\{1,2,3\}$ or $\{4,5,6\}$, e.g.\ for $k=2$ one has $j=1$ and $i=3$; and $\{l,m,n\}=\{1,\dots,6\}\backslash\{i,j,k\}$. 
Once the $d_{k}^{\pm}$ are expressed in terms of the cross ratios $\chi_{ab}$, they become independent of $\chi_{14}$.
We choose a basis of 9 last entries given by $w_{0}$, $w_{\chi_{14}}$, $\tau$ and the 6 torus images $w_{c_k}=w_{d_{k}^+}+w_{d_{k}^-}\,\text{mod}\,\tau$, where
\begin{align}
c_k=\chi_{14}\,\frac{\mathcal{G}_i\,x_{ik}^4+\mathcal{G}_j\,x_{jk}^4+2(\mathcal{G}_j^i+\mathcal{G}_{ij}\,x_{ij}^2)x_{ik}^2x_{jk}^2}{2\,\mathcal G_{ij}\,x_{ij}^2x_{jk}^2x_{ik}^2}~.
\label{eq:ci}
\end{align}
This basis spans the same space as the last entries obtained by the Schubert analysis.

\section{Bootstrap and Results}
\label{sec: results}

\begin{table}
\begin{center}
\begin{tabular}{l|l}
Constraint						& Free parameters\\ 
\hline\vspace{-10pt}
Alphabet					& 9 $\times$ 39 $\times$ 134 $\times$ 8\\
								& $\underbrace{\hspace{27pt}}$\\\vspace{-10pt}
Integrability in slot 1 \& 2  	& \hspace{9pt}60\hspace{12pt}$\times$ 134 $\times$ 8 \\ 
								& \hspace{41pt}$\underbrace{\hspace{32pt}}$\\\vspace{-10pt} 
Integrability in slot 3 \& 4  	& \hspace{9pt}60\hspace{12pt}$\times$\hspace{14pt}19\\
								& \hspace{9pt}$\underbrace{\hspace{52pt}}$\\
Integrability in slot 2 \& 3  	& \hspace{33pt}1 \\\vspace{-8pt} \\
Differential equation \eqref{eq: differential equation}	& \hspace{33pt}0
\end{tabular}
\caption{Number of remaining free parameters after imposing each constraint. 
}
\label{tab:dimensions}
\end{center}
\end{table}

Let us now turn to the bootstrap of the twelve-point double-box symbol assuming the structure \eqref{eq: symbol structure}; i.e.\ based on the alphabet generated in Sec.\ \ref{sec: schubert}, we make an ansatz for the terms in the symbol whose last entry is not $2\pi i\,\tau$, while we assume that the terms with last entry $2\pi i\,\tau$ follow from those via eq.\ \eqref{eq:intrep_symbolprime}.

A generic symbol $\sum_{i_1,\dots,i_n} C^{i_1, \dots, i_n} A_{i_1} \otimes \dots \otimes  A_{i_n}$ does not correspond to the symbol of a function unless it satisfies the integrability condition~\cite{Chen:1977oja}
\begin{align}
\label{eq:integrability_condition}
0 = & \, \sum_{i_1,\dots,i_n} C^{i_1, \dots, i_n} A_{i_1} \otimes \dots \otimes A_{i_{p-1}} \otimes A_{i_{p+2}} \otimes \dots \otimes A_{i_n} \nonumber \\
& \, \times \Big( \frac{\partial A_{i_{p}}}{\partial X_k} \, \frac{\partial A_{i_{p+1}}}{\partial X_m}-\frac{\partial A_{i_{p}}}{\partial X_m} \, \frac{\partial A_{i_{p+1}}}{\partial X_k} \Big)
\end{align}
at all depths $1\leq p<n$, where 
$\{X_{k}\}$ are a set of independent kinematic parameters, e.g.\ $\{\chi_{ab}\}$ or $\{w_0,w_{\chi_{14}},w_{c_k},\tau\}$ for the double-box integral. In particular, in order to exploit the structure \eqref{eq: symbol structure}, we use the latter set of variables 
for integrability in entries three and four; 
see Apps.\ \ref{sec: integrability} and \ref{sec: appendix_symbol_prime} for more details.

Amazingly, we find that imposing integrability uniquely fixes the symbol up to an overall constant, cf.\ Tab.\ \ref{tab:dimensions}!
We determine this constant via the differential equation \eqref{eq: differential equation}, which moreover provides a consistency check. In addition, we checked that the symbol satisfies the conformal Ward identity \cite{Chicherin:2017bxc}, the second-order differential equation of refs.\ \cite{Drummond:2006rz,Drummond:2010cz}, and reduces to the known symbol of the ten-point double box \cite{Kristensson:2021ani} in the limit $x_{16}^2\to0, x_{34}^2\to0$.
It also satisfies the (extended) Steinmann conditions \cite{Steinmann,Steinmann2,Caron-Huot:2019bsq} in all logarithmic symbol entries, i.e.\ discontinuities in overlapping channels vanish.\footnote{We leave the investigation of extended Steinmann conditions for the final entries for future work.}
Finally, 
the dual diagram of the double box is invariant under the $\mathbb{Z}_{2}$ reflection $x_{i}\to x_{7-i}$ and the permutations $S_{3}$ of $\{x_{1},x_{2},x_{3}\}$, cf.\ Fig.\ \ref{fig: integral 12pt}, and this symmetry is manifest in our symbol result.\footnote{This symmetry is lost in eq.\ \eqref{db_integrand} due to the normalization factor in the numerator,
but recovered in $I_{\db}/\omega_{1}$ and hence in eq.~\eqref{eq: copofdb}. We discuss the manifestation of these symmetries at the level of the symbol in more detail in App.\ \ref{sec:symb211}.} 
 
 The full symbol of the twelve-point double-box integral can be written in terms of 100 logarithmic symbol letters and 9 elliptic last letters, together with the structure shown in eq.~\eqref{eq:intrep_symbolprime}. We give its explicit form, as obtained from the bootstrap and organised by the last entries, in App.\ \ref{sec:symb211} as well as in an ancillary file.

Reorganizing this symbol allows to write the $(2,2)$-coproduct 
 of the double-box integral as a remarkably compact formula: 
\begin{align} 
&\Delta_{2,2}\bigg(\frac{2\pi i}{\omega_1}I_{\db}\bigg) \label{eq: copofdb}\\
&=\sum_{i<j}\text{Box}_{ij}\otimes \frac{2\pi i}{\omega_{1}} \int_{-\infty}^{\chi_{14}}\frac{d\chi_{14}'\log R_{ij}(\chi_{14}') }{\sqrt{-\Delta_{6}(\chi_{14}')}} 
 - (\chi_{14}{\to} \infty)\,, \nonumber
\end{align}
where the limit in the second term is taken with all other eight $\chi_{ab}$'s fixed.
The first term in eq.\ \eqref{eq: copofdb} manifests the differential equation \eqref{eq: differential equation} via eq.\ \eqref{eq: hexagon symbol}, and the subtracted term ensures integrability and
that $I_{\db}$ vanishes as $\chi_{14} \to \infty$.
The explicit form of the subtracted terms as a tensor product of weight-two functions can be understood as follows. Its first (weight-two) entries can be easily obtained by taking the limit $\chi_{14} \to \infty$ in the four-mass box terms \eqref{eq:Box}, e.g.\ $\text{Box}_{36}\to \mathcal{S}(\log x_{15;24} \log x_{14;25})$.
In order to obtain the second (weight-two) entries, we need to carefully define the integration contour connecting the two endpoints at infinity. This contour does not follow from the bootstrap, but is connected to the one in eq.~\eqref{eq:intrep_symbolprime}; we leave its investigation to future work.

The symbol of the second (weight-two) entry in the $(2,2)$-coproduct \eqref{eq: copofdb} can be written as
\begin{align} \label{eq: Sij_def}
\hspace{-0pt}S_{ij}&=\mathcal{S}\biggl(\frac{2\pi i}{\omega_{1}}\int\frac{d\chi_{14}'\log R_{ij}(\chi_{14}')}{\sqrt{-\Delta_{6}(\chi_{14}')}}\biggr)
=\hat{S}_{ij}+\mathbf{\Omega}_{ij}\otimes 2\pi i\, \tau,
\end{align}
where $\mathbf{\Omega}_{ij}$ can be obtained by taking the $\tau$-derivative of the integral in eq.\ \eqref{eq: Sij_def}, which is nothing but a realization of eq.\ \eqref{eq:intrep_symbolprime}.
The $\hat{S}_{ij}$ are given as follows:
If $i$ and $j$ are both taken from either $\{1,2,3\}$ or $\{4,5,6\}$, then let us denote $k$ as the respective third index from this set (with $\{i,j,k\}$ in cyclical ordering) and the associated symbol is
\begin{align}
&\hspace{-4pt}\hat{S}_{ij}=\log R_{ij}\otimes 2\pi i\,w_{\chi_{14}}+\frac 12\log\frac{\mathcal G_jx_{jk}^4}{\mathcal G_ix_{ik}^4}\otimes 2\pi i\,w_{c_{k}}\label{eq: Sij equal}\\
&\hspace{10pt}-\frac{1}{2}\!\sum_{l\in\{i,j\}}\!\!\!\text{sgn}(k-l)\log\frac{\mathcal G_{ijk\backslash l}^{ij}-\sqrt{\mathcal G_{ij}\mathcal G_{ijk\backslash l}}}{\mathcal G_{ijk\backslash l}^{ij}+\sqrt{\mathcal G_{ij}\mathcal G_{ijk\backslash l}}}\otimes 2\pi i\, w_{c_l}.\nonumber
\end{align}
Here, $\mathcal G_{ijk\backslash l}\equiv\mathcal G_{ik}$ if $l=j$ and $\mathcal G_{ijk\backslash l}\equiv\mathcal G_{jk}$ if $l=i$. 
If instead $i$ and $j$ take one value from each set,
e.g.\ $i\in\{1,2,3\}$ and $j\in\{4,5,6\}$,
then
\begin{align}
\hat{S}_{ij}=&\,\log R_{ij}\otimes 2\pi i\,w_{\chi_{14}}
\label{eq: Sij unequal}\\
&+\frac{(-1)^{i+j}}2\log\frac{z^2_{ij}(1-\bar{z}_{ij})}{\bar z^2_{ij}(1-z_{ij})}\otimes 2\pi i\,w_{0}\nonumber\\
&+\hspace{-4pt}\sum_{l\notin\{i,j\}}\hspace{-6pt}\frac{(-1)^l}2\log\frac{\mathcal G_{mn}^{ij}-\sqrt{\mathcal G_{ij}\mathcal G_{mn}}}{\mathcal G_{mn}^{ij}+\sqrt{\mathcal G_{ij}\mathcal G_{mn}}}\otimes 2\pi i\, w_{c_{l}}\,,\nonumber
\end{align}
where $m$ and $n$ are defined from $l$ by identifying the set $\{l,m,n\}$ with (cyclic permutations) of $\{1,2,3\}$ or $\{4,5,6\}$.
When taking the limit $\chi_{14}\to\infty$ to determine the subtracted term in eq.\ \eqref{eq: copofdb}, the 6 symbols \eqref{eq: Sij equal} vanish while the 9 symbols \eqref{eq: Sij unequal} yield 4 linearly independent combinations, resulting in the 19 integrable combinations found in Tab.\ \ref{tab:dimensions}.

\section{Conclusion and Outlook}
\label{sec: conclusion}

 In this letter, we have initiated the symbol bootstrap for elliptic Feynman integrals. 
 Concretely, we have determined the symbol of the two-loop twelve-point double-box integral.
 This calculation made use of two crucial ingredients: the simple structure \eqref{eq: symbol structure} of the symbol in terms of the symbol prime and a Schubert analysis to predict the symbol letters. 
 In particular, we show for the first time how a Schubert analysis can be used also to predict elliptic symbol letters.
Amazingly, our assumptions on the symbol alphabet combined with integrability were sufficient to uniquely determine the result up to an overall normalization, which we could fix via the differential equation \eqref{eq: differential equation}!
Moreover, we found a very compact expression for the $(2,2)$-coproduct, which in particular suggests that symbol-level integration \cite{CaronHuot:2011kk}
 can be generalized to the elliptic case.
 Note that while we have focussed on the symbol, it is also straightforward to complete the symbol by boundary values at a basepoint to a form that allows for numeric evaluation.
 
 We expect that the techniques developed in this work can be used to determine many further Feynman integrals and scattering amplitudes.
 A particular target would be all planar two-loop amplitudes in $\mathcal{N}=4$ sYM theory, which can be expressed in terms of a finite basis of elliptic Feynman integrals using prescriptive unitarity \cite{Bourjaily:2021vyj}.
Moreover, it would be interesting to make contact with the diagrammatic co-action \cite{Abreu:2017enx,Abreu:2021vhb} and spherical contours~\cite{Abreu:2017ptx,Arkani-Hamed:2017ahv}.
Many elliptic integrals that are relevant for LHC phenomenology contain massive internal propagators. It would be desirable to generalize the bootstrap approach and Schubert analysis also to this case. 
Finally, it would be very interesting to generalize the techniques developed here for elliptic integrals also to Feynman integrals containing higher-dimensional Calabi-Yau manifolds \cite{Bourjaily:2018ycu,Bourjaily:2018yfy,Bourjaily:2019hmc,Vergu:2020uur,Bonisch:2021yfw,Bourjaily:2022bwx,Duhr:2022pch,Pogel:2022ken,mirrors_and_sunsets}.

\begin{acknowledgments}

We thank 
Nima Arkani-Hamed,
Simon Caron-Huot, 
 James Drummond,
 Claude Duhr,
 Song He, Andrew McLeod,
 Marcus Spradlin,
 Cristian Vergu, Matt von Hippel, and
 Stefan Weinzierl
for fruitful discussions, as well as
Ruth Britto,
Simon Caron-Huot, 
Lance Dixon, and 
Marcus Spradlin
for comments on the manuscript.
QY and CZ are grateful to Song He for sharing ideas and collaborations on related projects.
The work of RM, AS, MW and CZ was supported by the research grant  00025445 from Villum Fonden and the ERC starting grant 757978. AS has furthermore received funding from the European Union's Horizon 2020 research and innovation program under the Marie Sk\l odowska-Curie grant agreement No.\ 847523 `INTERACTIONS'.

\end{acknowledgments}

\begin{appendix}

\section{Last entries from Schubert analysis}
\label{sec: Schubert appendix}

In this appendix, we elaborate on the procedure of obtaining the last entries of Sec.\ \ref{sec: schubert}. Following the argument in ref.\ \cite{Vergu:2020uur} and the discussion of Fig.\ \ref{fig: Schubert elliptic} in Sec.\ \ref{sec: schubert}, the leading singularity of the twelve-point double-box integral is
\begin{equation} \label{eq: LS_db}
    \operatorname{LS} I_{\db} = \operatorname{Res}_{Q_{L}}
     \operatorname{Res}_{Q_{R}} \frac{ \langle (1)(4)\rangle\langle (2)(5)\rangle \langle (3)(6)\rangle\langle Xd^{3}X\rangle}{\langle X(1)(2)(3)\rangle \langle X(4)(5)(6)\rangle} \:,
\end{equation}
where $\langle Xd^{3}X\rangle = \epsilon^{\mu\nu\rho\sigma}X_{\mu}dX_{\nu}dX_{\rho}dX_{\sigma}$ is the $\rm{PSL}$(4)-invariant top form in $\mathbb{P}^{3}$. Here $\operatorname{Res}_{Q_{L,R}}\omega$ means taking the residue of the differential form $\omega$ on the quadratic surface $Q_{L}$ or $Q_{R}$ given by
\begin{align} \label{eq: QLQR}
    Q_{L,R}\,:\quad 0 &=\langle X(i)(j)(k)\rangle  \\
	& \equiv 
    \langle X A_{i}B_{i}A_{k}\rangle \langle XA_{j}B_{j}B_{k}\rangle  - (A_{k}\leftrightarrow B_{k}) \:, \nonumber  
\end{align}
where $A_{i}$ and $B_{i}$ are two distinct points on the line $(i)$ defined by eq.~\eqref{eq: line_def}, such that the factor $\langle (i)(j)\rangle$ in the numerator of eq.~\eqref{eq: LS_db} is simply $\langle A_{i}B_{i} A_{j}B_{j}\rangle$. For our purpose, we can simply parametrize $X$ as $  X= Z_{\alpha} + x Z_{\beta} + \gamma Z_{A_{j}} + \delta Z_{B_{j}}$, where $\alpha$ and $\beta$ are intersection points of the line $(i)$ with the two solution lines of the one-loop Schubert problem containing either $\{(1),(2),(3)\}$ or $\{(4),(5),(6)\}$. Solving $\gamma$ and $\delta$ in terms of $x$ in eq.~\eqref{eq: LS_db} through eq.~\eqref{eq: QLQR} leads to $\operatorname{LS} I_{\db} = dx/\sqrt{P(x)}$,
where $P(x)$ is a \emph{quartic} polynomial in $x$ since it arises from the intersection of two quadratic surfaces in $\mathbb{P}^{3}$. Since here we choose to parametrize the line $(i)$ with $\alpha$ and $\beta$,  eq.~\eqref{eq: elllast} becomes
\begin{equation}
    \frac{2\pi i}{\omega_{1}} \int_{0}^{\infty} \frac{dx}{\sqrt{P(x)}} \:.
\end{equation}
Note that there are two homotopically inequivalent (up to $A$- and $B$-cycles) choices for the contour connecting $0$ and $\infty$, see Fig.\ \ref{fig: contours}. Taking both cases into account gives the full set of last entries.

\tikzset{cross/.style={cross out, thick, draw=black, fill=none, minimum size=2*(#1-\pgflinewidth), inner sep=0pt, outer sep=0pt}, cross/.default={2pt}}
\begin{figure}
	\centering
	\begin{tikzpicture}
			\draw[->, thick] (-1,0) to (-1,3);
			\draw[->, thick] (-3,1.5) to (1,1.5);
			\draw[line width=0.3mm,red] (-1,1.5) to (0.7,1.5);
			\draw[line width=0.3mm,red] (-1,1.5) to (-1,0.2);
			\node[cross,label=above:$r_{1}$] at (-1.5, 2.0) {};
			\node[cross,label=below:$r_{4}$] at (-1.5, 1.0) {};
			\node[cross,label=above:$r_{2}$] at (-0.5, 0.8) {};
			\node[cross,label=below:$r_{3}$] at (-0.5, 2.2) {};
			\node at (-1.2,1.7) {$O$};
			\node at (0.7,1.7) {$\infty$};
			\node at (-1.3,0.2) {$\infty$};
			\draw[fill=ku] (-1,1.5) circle (2pt);
			\draw[fill=ku] (0.7,1.5) circle (2pt);
			\draw[fill=ku] (-1,0.2) circle (2pt);
			\node at (-0.7,3) {$\Im x$};
			\node at (1.3,1.5) {$\Re x$};
	\end{tikzpicture}
	\caption{The two homotopically inequivalent contours (in red) connecting $0$ and $\infty$ (up to $A$- and $B$-cycles). Here we assume that the four roots of $P(x)$ fall into four quadrants, without loss of generality. 
	}
	\label{fig: contours}
	\end{figure}
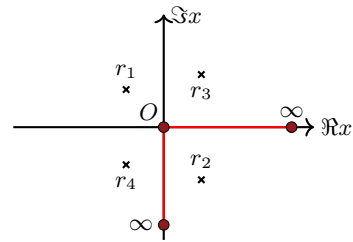

\section{Constructing spaces via integrability}
\label{sec: integrability}
In this appendix, we describe how to apply the integrability conditions \eqref{eq:integrability_condition} to elliptic letters in order to constrain the alphabet and the symbol of elliptic Feynman integrals.

In our case, the last entries of the symbol of the double box are   8 elliptic letters $w_{\notc}$ and $\tau$, where the precise values of $\notc$ are constructed in Sect.\ \ref{sec: schubert}. 
In order to exploit the structure \eqref{eq: symbol structure} of the symbol, we want to calculate derivatives precisely with respect to these variables when imposing integrability in entries three and four, i.e.\ we take $X_k \in \{ w_{\notc_1},\dots,w_{\notc_8},\tau \}$ in eq.~\eqref{eq:integrability_condition}. Then, derivatives on the last entries act trivially, $\partial w_{\notc_i}/\partial w_{\notc_j}=\delta_{ij}$ and $\partial\tau/\partial w_{\notc_j}=0$. 
As discussed in App.\ \ref{sec: appendix_symbol_prime}, the integrability for last entry being $\tau$ is granted by the symbol prime. Thus, we will only encounter derivatives of the type 
$\partial \log(\phi_i(\chi_{ab}))/\partial w_\notc$ for the next-to-last entries, where $\phi_i(\chi_{ab})$ is a rational or algebraic function of the 9 independent dual-conformal cross ratios.
To calculate these derivatives via the chain rule, we require 
the 
elements of the inverse Jacobian $J^{-1}=\frac{\partial (\chi_{a_1 b_1}, \dots, \chi_{a_n b_n})}{\partial (w_{\notc_1}, \dots, w_{\notc_{n-1}}, \tau)}$, where we have generalized to $n+1$ variables.
Starting from Abel's map \eqref{eq:Abel_map} and using the family of integrals and recursion relations provided in ref.\ \cite{Broedel:2017kkb}, the elements of the Jacobian $J$ become
\begin{align}
\label{eq:integrability_cubic_wc}
\frac{\partial w_\notc}{\partial \chi_{ab}} = & \, \frac{1}{\omega_1 \bar{y}_\notc} \frac{\partial \bar{\notc}}{\partial \chi_{ab}} + \frac{1}{\omega_1} \sum_{k=1}^3 \frac{\bar{y}_\notc}{(\bar{r}_k - \bar{\notc} ) \prod_{j \neq k} (\bar{r}_k - \bar{r}_j)} \frac{\partial \bar{r}_k}{\partial \chi_{ab}} \nonumber \\
& \, - \frac{2}{\omega_1^2} g^{(1)}(w_{\bar{\notc}}) \sum_{k=1}^3 \frac{1}{\prod_{j \neq k} (\bar{r}_k - \bar{r}_j)} \frac{\partial \bar{r}_k}{\partial \chi_{ab}}, \\
\label{eq:integrability_cubic_tau}
\frac{\partial \tau}{\partial \chi_{ab}} = & \, \frac{4 \pi i}{\omega_1^2} \sum_{k=1}^3 \frac{1}{\prod_{j \neq k} (\bar{r}_k - \bar{r}_j)} \frac{\partial \bar{r}_k}{\partial \chi_{ab}},
\end{align}
where the overline is a short-hand notation for $\bar{x} \equiv a_3 x$, with $a_3$ being the coefficient of the cubic term in the elliptic curve. In addition, $y_\notc$ are the images on the elliptic curve, $r_i$ are its roots, and $g^{(1)}(w)=\partial_w \log{\theta_1(w)}$ for $\theta_1(w)$ being the odd Jacobi theta function.

For completeness, let us note that the corresponding expressions for an elliptic curve given by a monic quartic polynomial similarly become
\begin{align}
\label{eq:integrability_quartic_wc}
\frac{\partial w_\notc}{\partial \chi_{ab}} = & \, \frac{1}{\omega_1 y_\notc} \frac{\partial \notc}{\partial \chi_{ab}} + \frac{1}{\omega_1} \sum_{k=1}^4 \frac{y_\notc + (r_k - \notc )^2}{(r_k - \notc ) \prod_{j \neq k} (r_k - r_j)} \frac{\partial r_k}{\partial \chi_{ab}} \nonumber \\
& \, - \frac{2}{\omega_1^2} g^{(1)}(w_\notc) \sum_{k=1}^4 \frac{1}{\prod_{j \neq k} (r_k - r_j)} \frac{\partial r_k}{\partial \chi_{ab}}, \\
\label{eq:integrability_quartic_tau}
\frac{\partial \tau}{\partial \chi_{ab}} = & \, \frac{4 \pi i}{\omega_1^2} \sum_{k=1}^4 \frac{1}{\prod_{j \neq k} (r_k - r_j)} \frac{\partial r_k}{\partial \chi_{ab}}.
\end{align}

In both cases, the functions $g^{(1)}(w)$ drop in the elements $\partial \chi_{ab}/\partial w_\notc$ of the inverse Jacobian and thus also drop in the integrability conditions for the last entries being $w_\notc$, rendering these conditions purely algebraic.

\section{Integrability and the symbol prime}
\label{sec: appendix_symbol_prime}

Integrability in the last two entries of eq.\ \eqref{eq: symbol structure} is satisfied if 
\begin{align}
\partial_{w_{\notc_a}}\!\!\log(\phi_{ib})-\partial_{w_{\notc_b}}\!\!\log(\phi_{ia})&=0\label{eq:lastpairint}~,\\
\partial_{w_{\notc_a}}\mathbf{\Omega}_i-\partial_\tau\log(\phi_{ia})&=0~.
\label{eq:symbolprimeint}
\end{align}
In our approach, the first condition implies an integrability constraint that we impose to fix the coefficients in our ansatz, while the second condition is manifest from the construction \eqref{eq:intrep_symbolprime} via the symbol prime, cf.\ the discussion in ref.\ \cite{Wilhelm:2022wow}. 

In this appendix, we study the integrability condition in the second-to-last entry pair, i.e.\ in the second and third entry for the double-box symbol. When using the symbol-prime construction, the last three entries of a symbol are generally of the form
\begin{align}
\log(\phi_l)\otimes\biggl[\sum_j\log(\phi_{ij})\otimes 2\pi i\, w_{\notc_j}+\mathbf{\Omega}_i\otimes 2\pi i\,\tau\biggr].
\end{align}
They satisfy the integrability conditions if
\begin{align}
\big[\partial_{w_{\notc_a}},\partial_{w_{\notc_b}}\big]\log(\phi_l)\otimes\log(\phi_{ij})&=0~,\\
\big[\partial_{w_{\notc_a}},\partial_\tau\big]\log(\phi_l)\otimes\log(\phi_{ij})&=0~,\label{eq:int2cond}\\
\big[\partial_{w_{\notc_a}},\partial_{w_{\notc_b}}\big]\log(\phi_l)\otimes\mathbf{\Omega}_i&=0~,\label{eq:yes}\\
\big[\partial_{w_{\notc_a}},\partial_\tau\big]\log(\phi_l)\otimes\mathbf{\Omega}_i&=0~, \label{eq:no}
\end{align}
where we again use the set of last entries $\{w_{\notc_a},\tau\}$ as the independent set of kinematic variables. The first and second conditions are manifest after imposing integrability in the bootstrapped part of the symbol. The left-hand side of the third condition is, using eq.~\eqref{eq:symbolprimeint}, equivalent to
\begin{align}
\big(\partial_{w_{\notc_a}}\!\!\log\phi_l\big)\big(\partial_\tau\log\phi_{ib}\big)-\big(\partial_{w_{\notc_b}}\!\!\log\phi_l\big)\big(\partial_\tau\log\phi_{ia}\big)~.
\end{align}
Now using eq.~\eqref{eq:int2cond} in both terms gives
\begin{align}
\big(\partial_\tau\log\phi_l\big)\big(\partial_{w_{\notc_a}}\!\!\log\phi_{ib}\big)-\big(\partial_\tau\log\phi_l\big)\big(\partial_{w_{\notc_b}}\!\!\log\phi_{ia}\big)~,
\end{align}
which vanishes according to eq.~\eqref{eq:lastpairint}, and thus integrability condition \eqref{eq:yes} is also manifest in our construction. The remaining constraint \eqref{eq:no} is non-trivial and will be further discussed in future work.

\section{Explicit double-box symbol}
\label{sec:symb211}

In this appendix, we give the symbol of the elliptic double-box integral organized by its nine elliptic last letters.
It is
\begin{align}
&\mathcal{S}\bigg(\frac{2\pi i}{\omega_1}I_{\db}\bigg)=\ \mathcal{S}(I_{\hex})\otimes(2\pi i\, w_{\chi_{14}})+F_\tau\otimes(2\pi i\,\tau)\nonumber\\
&+\tfrac 12\hspace{-7pt}\sum_{\substack{i\in\{1,2,3\}\\j\in\{4,5,6\}}}\hspace{-9pt}\text{V}_{ij}\otimes(2\pi i\,w_0)+\tfrac 12\hspace{-7pt}\sum_{k\in\{1,...,6\}}\hspace{-9pt}\text{W}_{k}\otimes(2\pi i\,w_{c_{k}})~.
\label{eq:finalS}
\end{align}
Here we introduced the definition
\begin{align}
\text{V}_{ij}=&\ (-1)^{i+j}\Big(\text{Box}_{ij}\otimes\log\frac{z_{ij}^2}{{\bar z}_{ij}^2}\frac{1-\bar{z}_{ij}}{1-z_{ij}}-\text{U}_{ij}\otimes \log v_{ij}\Big),
\label{eq:Vij}
\end{align}
where
\begin{align}
\text{U}_{ij}=
\mathcal S(\log(u_{ij})\log(v_{ij}/u_{ij}))
=\lim_{\chi_{14}\rightarrow\infty}\text{Box}_{ij}~.
\end{align}
Moreover, $\lim_{\chi_{14}\rightarrow\infty}\log\frac{z_{ij}^2}{{\bar z}_{ij}^2}\frac{1-\bar{z}_{ij}}{1-z_{ij}}=\log v_{ij}$ and thus eq.~\eqref{eq:Vij} mimics the structure in eq.~\eqref{eq: copofdb}. The last term in eq.~\eqref{eq:finalS} contains the weight-three symbols
\begin{align}
&\text{W}_k=\text{Box}_{ij}\otimes\log\frac{\mathcal G_j\hspace{0.5pt}x_{jk}^4}{\mathcal G_i\hspace{0.5pt}x_{ik}^4}\nonumber\\
&+(-1)^{i-j}\hspace{-5pt}\sum_{\substack{l\in\{i,j\}\\ m\not\in\{i,j\}}}\hspace{-7pt}\text{sgn}(m-l)\text{Box}_{lm}\otimes\log\frac{\mathcal G_{lm}^{ij}-\sqrt{\mathcal G_{ij}\mathcal G_{lm}}}{\mathcal G_{lm}^{ij}+\sqrt{\mathcal G_{ij}\mathcal G_{lm}}}\nonumber\\
&+(-1)^{i+j+k}\hspace{-8pt}\sum_{l\not\in\{1,6,i,j,k\}}\hspace{-10pt}(-1)^l\bigr(\text{U}_{il}-\text{U}_{jl}\bigl)\otimes\log\frac{1-z_{ij}}{1-\bar z_{ij}}\nonumber\\
&-(-1)^{i+j+k}\hspace{-8pt}\sum_{l\not\in\{3,4,i,j,k\}}\hspace{-10pt}(-1)^l\bigr(\text{U}_{il}-\text{U}_{jl}\bigl)\otimes\log\frac{z_{ij}}{\bar z_{ij}}~,\label{eq:Wi}
\end{align}
where the indices $i$ and $j$ are again defined from $k$ by identifying $\{i,j,k\}$ with (cyclic permutations of) $\{1,2,3\}$ or $\{4,5,6\}$. In the limit $\chi_{14}\rightarrow\infty$, the first term vanishes, as well as the terms with $m=k$ in the sum in the second line. The remaining terms in the sum, with $m\neq k$, reduce to the expressions in the last two lines of eq.\  \eqref{eq:Wi} up to a sign, i.e.\ also here the structure in eq.~\eqref{eq: copofdb} is manifest.

The weight-three symbols V$_{ij}$ and W$_k$ individually satisfy integrability and the (extended) Steinmann conditions \cite{Steinmann,Steinmann2,Caron-Huot:2019bsq}. $F_\tau$ is determined via the structure~\eqref{eq: symbol structure} up to a $\tau$-dependent contribution, cf.\ the discussion in Sec.~\ref{sec: results}. The only term in eq.\ \eqref{eq:finalS} with a last entry depending on $\chi_{14}$ is $\mathcal{S}(I_{\hex})\otimes(2\pi i\, w_{\chi_{14}})$, and thus the differential equation \eqref{eq: differential equation} is manifest ($\partial_{\chi_{14}}w_{\chi_{14}}=(\omega_1\sqrt{-\Delta_6})^{-1}$). 

Without the numerator, $I_{\db}$ is invariant under the $\mathbb{Z}_2$ transformation $x_i\to x_{7-i}$,  as well as under all permutations $S_3$ of $x_1$, $x_2$, $x_3$. With the numerator, $I_{\db}$ is only invariant under $\mathbb{Z}_2\times \mathbb{Z}_2\subset  \mathbb{Z}_2\times S_3$, where the second $\mathbb{Z}_2$ is generated by $x_1\leftrightarrow x_3$, $x_4\leftrightarrow x_6$, while it transforms covariantly under the remaining generators.
Under the left--right reflection symmetry $\mathbb{Z}_2$ of $I_{\db}$, the last entries transform via $w_{\chi_{14}}\rightarrow w_{\chi_{14}}$, $w_0\rightarrow w_0$, and $w_{c_{k}}\rightarrow-w_{c_{7-k}}$.  
Under each permutation of two dual points from either $S_3$ symmetry, $w_{\chi_{14}}$ stays invariant, whereas $w_0$ transforms to $-w_0$ (due to $y_0\rightarrow -y_0$). Under a permutation of, for example, the two dual points $x_5$ and $x_6$, the entries $w_{c_k}$ stay unchanged for $k\in\{1,2,3\}$, while $w_{c_4}$  picks up a sign, and $w_{c_5}\leftrightarrow-w_{c_6}$. These transformations are mimicked by their preceding weight-three symbols in eq.\ \eqref{eq:finalS}. The term $F_\tau\otimes 2\pi i\tau$, which is constructed from the remaining terms in the symbol, inherits these symmetry properties, such that whole symbol is invariant under $S_3\times \mathbb{Z}_2$.

\end{appendix}

\bibliography{reference}

\end{document}